\documentclass[aps,prc,preprint,tightenlines,showpacs,%
amssymb,byrevtex,nofootinbib]{revtex4}  

\usepackage{epsfig,bm,dcolumn}

\begin{document}

\preprint{hep-ph/0310019}


\title{Exotic $\Theta^+$ baryon production induced by photon and pion}


\author{Yongseok Oh}%
\email{yoh@phya.yonsei.ac.kr}

\author{Hungchong Kim}%
\email{hung@phya.yonsei.ac.kr}

\author{Su Houng Lee}%
\email{suhoung@phya.yonsei.ac.kr}

\affiliation{Institute of Physics and Applied Physics,
Yonsei University, Seoul 120-749, Korea}



\begin{abstract}

We investigate the photoproduction of the $\Theta^+(1540)$ on a nucleon
($\gamma n \to K^-  \Theta^+$, $\gamma p \to \bar{K}^0 \Theta^+$)
and the pion-induced $\Theta^+$ production reaction on the proton
($\pi^- p \to K^- \Theta^+$).
The total cross sections near threshold are estimated by using hadronic
models with effective interaction Lagrangians and form factors that preserve
gauge-invariance of the electromagnetic current.
The photoproduction cross sections are found to be a few hundred nb, with
the cross section on the proton being larger than that on the neutron.
The pion-induced production cross section is found to be around a few
hundred $\mu$b but sensitive to the $K^* N \Theta$ coupling whose value is
not yet known.
We also study the production cross section assuming that the $\Theta^+$ has
negative parity.
The cross sections are then found to be much suppressed compared to the
case where $\Theta^+$ has positive parity.
Hence, the interpretation of the $\Theta^+$ as an odd-parity pentaquark
state seems to be disfavored from the estimates of cross section for the
photon-proton reaction from the SAPHIR experiment.

\end{abstract}


\maketitle

\section{Introduction}

The recent interests in pentaquark exotic hadrons were triggered by the
discovery of the $\Theta^+(1540)$ baryon by the LEPS Collaboration at
SPring-8 \cite{LEPS03}, where the photon beam was used on a $^{12}$C
target to produce the pentaquark $\Theta^+$ from $\gamma n \to K^- \Theta^+$
reaction.
The upper limit of its decay width ($\Gamma_{\Theta}$) was estimated to
be 25 MeV.
The CLAS Collaboration at Thomas Jefferson National Accelerator Facility
used the photon-deuteron reaction to produce the $\Theta^+$ and found
the decay width to be less than 21 MeV \cite{CLAS03-b}.
The SAPHIR Collaboration used the photon-proton reaction
($\gamma p \to \bar{K}^0 \Theta^+$), where the decay width is found to be
less than 25 MeV \cite{SAPHIR03}.
The $\Theta^+$ production with $\Gamma_{\Theta} \le 9$ MeV was also reported
in the kaon-neutron reaction ($K^+ n \to K^0 p$) by the DIANA Collaboration
\cite{DIANA03}.
Recently the $\nu N$ reaction was used to search for the $\Theta^+$ with
$\Gamma_\Theta < 20$ MeV \cite{ADK03}.

Although the quantum numbers of the $\Theta^+(1540)$ are still to be
determined, the interpretation of it as being a pentaquark
$(uudd\bar{s})$ state is solid because the $\Theta^+$ has positive
strangeness ($S=+1$). 
Such a low-lying pentaquark state with narrow width was first predicted
in the chiral quark soliton model \cite{DPP97}, although the existence
of such exotic states was anticipated earlier in the study of the Skyrme
model \cite{Chem85,Weig98,Pras03}.
The recent experimental findings prompted a lot of theoretical reinvestigation
of the pentaquark states including the pentaquark $(P_{\bar Q})$ with
one heavy antiquark \cite{Lip87-GSR87,RS93,OPM94b-OPM94c,OP95,%
E791-98-E791-99, KL03b,JW03,Cheung03}.
Subsequent theoretical investigations on the $\Theta^+$ include approaches
based on the constituent quark model
\cite{SR03,CCKN03,Cheung03,Gloz03a-b,KL03a},
Skyrme model \cite{Weig98,WK03,Pras03,JM03,BFK03,IKOR03},
QCD sum rules \cite{Zhu03,MNNRL03,SDO03}, chiral potential model
\cite{Hosaka03}, large $N_c$ QCD \cite{Cohen03-CL03}, lattice QCD
\cite{CFKK03}, and Group theory approach \cite{Wyb03}.
The production of the $\Theta^+$ was also discussed in relativistic nuclear
collisions \cite{Rand03,CGKLL03}, where the number of the
anti-$\Theta^+(1540)$ produced are expected to be similar to that of the
$\Theta^+(1540)$. 
However the genuine structure of the $\Theta^+$ is still to be clarified,
e.g., it is not yet firmly established whether the $\Theta^+$ forms an
anti-decuplet with the Roper resonance \cite{Gloz03a-b}, and its
spin-parity is not yet confirmed.
On the other hand, Jaffe and Wilczek suggested diquark-diquark-antiquark
nature of the $\Theta^+$  in the anti-decuplet plus octet representation of
SU(3) \cite{JW03}.
In Ref. \cite{BM03}, the $\Theta^+$ is even claimed to be a heptaquark
state.
In addition, Capstick {\it et al.\/} suggested $\Theta^+$ as a member of
isotensor pentaquark family \cite{CPR03}, which is, however, doubted by the
SAPHIR experiment.

In the midst of such confusion, one attempt is to assume certain quantum
numbers for the $\Theta^+$ and investigate its physical properties
\cite{HHO03,PR03,CMM03,GM99}.
As a starting point to compare with experimental observations, it is
important to investigate the production processes of the $\Theta^+$ in
the photon-induced and pion-induced reactions.
Since the production processes are studied in the medium energy
region, hadronic description would be more appropriate than perturbative
QCD.
There have been studies in this direction, where a hadronic model with
effective interaction Lagrangians was used to calculate the reaction
cross sections.
In Refs. \cite{LK03a,LK03b}, Liu and Ko estimated the cross sections of
positive-parity $\Theta^+$ production from photon-nucleon scattering and
various meson-nucleon scatterings.
The authors considered not only the 2-body final states, but also 3-body
final states.
They claimed that the cross sections are about $0.05$ mb in pion-nucleon
reaction, $40$ nb in photon-proton reaction, and $280$ nb in
photon-neutron reaction \cite{LK03a}.
The cross section for photon-neutron reaction is claimed to be
substantially larger than that for photon-proton reaction.
The values are changed in their sequential work \cite{LK03b}, which
includes the contributions from the $K^*$ exchanges in the photon-nucleon
reactions.
However, this work does not take into account the tensor coupling of the
photon-nucleon and photon-$\Theta^+$ interactions.
In particular, they did not include the $s$-channel diagrams and the
anomalous magnetic momentum terms in the $u$-channels in photoproduction
reaction, which were shown to be important in Ref. \cite{NHK03}.
In addition, the final results were obtained by multiplying a form factor
which is a function of the center-of-mass energy only.
In Ref. \cite{NHK03}, Nam {\em et al.\/} considered $\gamma n \to K^-
\Theta^+$ process using pseudoscalar and pseudovector couplings as well
as a hybrid model.
Then the authors included the effects of the form factors by dividing the
cross section by an overall energy-independent constant, whose value is
obtained from a similar prescription to match the theoretical
Born term estimate of the total cross section for kaon photoproduction to
the experimental data.
They also considered the case where the quantum numbers of the
$\Theta^+$ are $J^P = \frac12^-$.
Then they found that the $\Theta^+$ production cross section in
photon-neutron reaction near threshold is 14--20 nb for negative-parity
$\Theta^+$ and 100--240 nb for the positive-parity $\Theta^+$.
However, in their work, the $K^*$ exchange was not considered and the
assumption that the $\Theta^+$ production cross section can just be
divided by a constant factor is unjustified.
Experimentally, the only information available for any production cross
sections comes from the SAPHIR Collaboration \cite{SAPHIR03}, which claims
that the cross section for $\gamma p \to \bar{K}^0 \Theta^+$ is similar to
that of $\phi$ photoproduction and is order of 200 nb near threshold,
which is to be confirmed by further analyses \cite{Barth}.

In this work, we perform a more consistent calculation on the
photoproduction of the $\Theta^+$ from the nucleon targets and on
the pion-induced production from the proton target.
The latter reaction is of particular interest since the current KEK
experiment searching for the $\Theta^+$ is using this reaction.
Such a reaction can also be studied with the recent pion beam facility
at GSI.
Several improvements are included in our work compared with previous
hadronic model calculations in Refs. \cite{LK03a,LK03b,NHK03}.
To investigate the sensitivity on the possible form factors, we employ
form factors that are functions of the transferred momenta and compare
the results with the previous ones that use different prescriptions for
the form factors.
We also include the $K^*$ exchanges in the $t$-channel in all relevant
reactions.
As we will show, the contributions from the $K^*$ exchange is
appreciable in all the production reactions considered and in fact
dominant in the pion-induced reaction.
Another important question that we address is the parity of the $\Theta^+$,
which is not yet settled.
For example, Refs. \cite{CCKN03,Zhu03,SDO03,CFKK03} suggest that the parity
of the $\Theta^+$ is preferably odd, while many other approaches including
soliton models claim or assume it to be even.
Therefore we will first present the results assuming that the
$\Theta^+(1540)$ is an isosinglet, spin-1/2 baryon with positive parity,
and then the results with assuming that the $\Theta^+$ has negative
parity will be compared and discussed.

\section{$\bm{\gamma n \to K^- \Theta^+}$ and
$\bm{\gamma p \to \bar{K}^0 \Theta^+}$}

\begin{figure}[t]
\centering
\epsfig{file=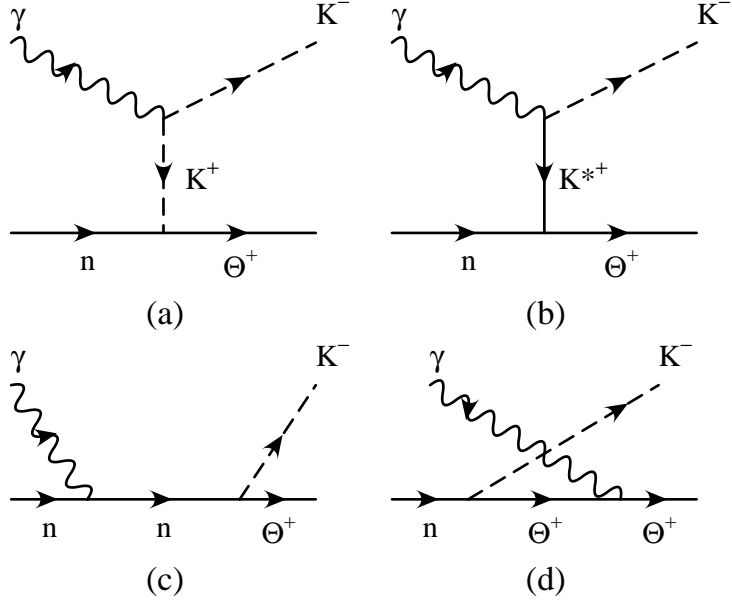, width=0.6\hsize}
\caption{Diagrams for $\gamma n \to K^- \Theta^+$ reaction.}
\label{fig:gam-n}
\end{figure}

\begin{figure}[t]
\centering
\epsfig{file=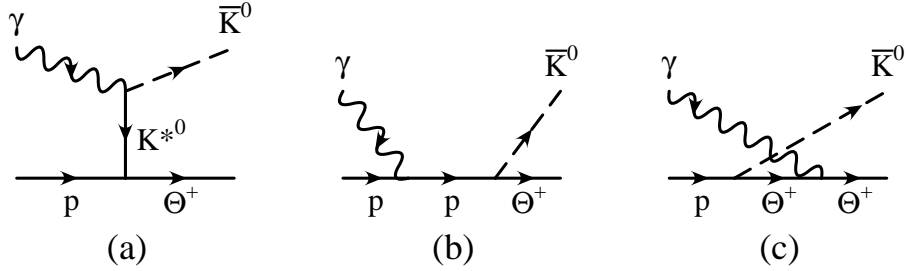, width=0.75\hsize}
\caption{Diagrams for $\gamma p \to \bar{K}^0 \Theta^+$ reaction.}
\label{fig:gam-p}
\end{figure}

The Feynman diagrams of $\Theta^+$ photoproduction from the neutron
and proton targets are shown in Figs. \ref{fig:gam-n} and
\ref{fig:gam-p}.
The momenta of the incoming photon, the nucleon, the outgoing $K$, and
the $\Theta$ are $k$, $p$, $q$, and $p'$, respectively.
The Mandelstam variables are $s=(k+p)^2$, $t=(k-q)^2$, and $u=(p-q)^2$.
It should be noted that we have neglected the $s$-channel diagrams in
which the intermediate baryon is the nucleon resonance, including the
Roper resonances or the non-strange analog of the $\Theta^+$ that could
be the Roper $N(1710)$ \cite{DPP97,JW03}.
Such approximations should be good enough in a first attempt
calculation, where at least all the ground state nucleon, pseudoscalar
and vector mesons are consistently included.

For the parity of the $\Theta^+$, we note that the chiral quark soliton
model predicts even parity.
This seems to be consistent with the Skyrme model results on the
pentaquark states containing one heavy anti-quark.
In that model, the lowest state of pentaquark with one heavy antiquark
and four light $u,d$ quarks has $I=0$ and $J^P = \frac12^+$,
while the first excited state has $I=0$ and $J^P = \frac12^-$,
and the $I=1$ pentaquarks are higher states \cite{OP95}.
The $\Theta^+$ with $J^P=\frac12^+$ is also favored by a recent Skyrme
model study \cite{IKOR03} and a constituent quark model study
\cite{CCKN03b}.
Thus, we first assume that the $\Theta^+$ has positive parity postponing
the negative-parity case to Sec. IV.
Then the effective Lagrangians read
\begin{eqnarray}
\mathcal{L}_{\gamma KK} &=& ie A_\mu ( K^- \partial^\mu K^+ - \partial^\mu
K^- K^+),
\nonumber \\
\mathcal{L}_{KN\Theta} &=& -i g_{KN\Theta}^{} ( \bar{\Theta} \gamma_5 K^+ n
- \bar{\Theta} \gamma_5 K^0 p ) + \mbox{ h.c.},
\nonumber \\
\mathcal{L}_{\gamma\Theta\Theta} &=& -e \bar{\Theta} \left[ A_\mu \gamma^\mu
- \frac{\kappa_\Theta^{}}{2M_\Theta} \sigma_{\mu\nu} \partial^\nu A^\mu
  \right] \Theta,
\nonumber \\
\mathcal{L}_{\gamma N N} &=& -e \bar{N} \left[ A_\mu \gamma^\mu
\frac{1+\tau_3}{2}
- \frac{1}{4M_N} \left\{ \kappa_p + \kappa_n + \tau_3 (\kappa_p -
  \kappa_n) \right\} \sigma_{\mu\nu} \partial^\nu A^\mu
  \right] N,
\label{Lag:Theta}
\end{eqnarray}
where $A_\mu$ is the photon field and $N^T = (p, n)$.
The anomalous magnetic moments of the proton and neutron are
$\kappa_p(=1.79)$ and $\kappa_n(=-1.91)$, respectively.
Here we use the SU(3) Lagrangian for the phases of the $\Theta^+ KN$
interactions \cite{OKL03b}.

For the $K^*$ exchange, we use
\begin{eqnarray}
\mathcal{L}_{K^* K \gamma} &=& g_{K^* K \gamma}^{}\,
\varepsilon^{\mu\nu\alpha\beta} \partial_\mu A_\nu \left(
\partial_\alpha K_\beta^{*-} K^+ + \partial_\alpha \bar{K}^{*0}_\beta
K^0 \right) + \mbox{ h.c.},
\nonumber \\
\mathcal{L}_{K^* N \Theta} &=& -g_{K^*N\Theta}^{}
\bar{\Theta} \left( \gamma^\mu K^{*+}_\mu - \frac{\kappa_{K^* n
\Theta}^{T}}{M_N+M_\Theta} \sigma^{\mu\nu} \partial_\nu K_\mu^{*+} \right)
n
\nonumber \\ && \mbox{}
+  g_{K^*N\Theta}^{}
\bar{\Theta} \left( \gamma^\mu K^{*0}_\mu - \frac{\kappa_{K^* p
\Theta}^{T}}{M_N+M_\Theta} \sigma^{\mu\nu} \partial_\nu K_\mu^{*0} \right) p
+ \mbox{ h.c.}.
\label{Lag:K*}
\end{eqnarray}

The coupling constants are determined as follows.
The Lagrangian $\mathcal{L}_{KN\Theta}$ gives the decay width of
$\Theta^+ \to KN$ as
\begin{equation}
\Gamma_{\Theta^+ \to K^+ n + K^0 p}
= \frac{g_{KN\Theta}^2}{2\pi} \frac{|{\bf p}_{K}^{}|
( \sqrt{M_N^2 + {\bf p}_K^2}-M_N)}{M_\Theta},
\label{decay}
\end{equation}
where $M_N$ and $M_\Theta$ are the nucleon and $\Theta^+$ mass,
respectively, and ${\bf p}_K^{}$ is the momentum of the kaon in the
$\Theta^+$ rest frame.
Thus, the coupling $g_{KN\Theta}^{}$ can be estimated from the
$\Theta^+$ decay width.
In Ref. \cite{CMM03}, the decay ratio $\Gamma_{\Theta \to K^+ n} /
\Gamma_{\Theta \to K^0 p}$ was shown to be dependent on the isospin of
the $\Theta^+$.
If the $\Theta^+$ is an isosinglet, this ratio becomes one.
Theoretically, the chiral soliton model of Ref. \cite{DPP97} predicted a
very narrow width of less than 15 MeV.
Later it was claimed to be about 5 MeV in an improved analysis of the
same model \cite{PSTCG00}.
This small decay width seems to be consistent with recent analyses on $KN$
scattering that suggest a narrow width of a few MeV for the $\Theta^+$
\cite{Nuss03,ASW03,CN03,HK03}.
Experimentally, only the upper bound of the $\Theta^+$ decay width is
known, around 9--25 MeV.
If we take the results of the chiral quark soliton model \cite{PSTCG00} and
the $KN$ scattering analyses \cite{Nuss03,ASW03,CN03,HK03}, which is
$\Gamma_{\Theta^+ \to K N} = 5 \sim 10$ MeV, we get
\begin{equation}
g_{KN\Theta}^{} = 2.2 \sim 3.11.
\end{equation}
This value is much smaller than $g_{KN\Lambda}^{}$, which is $-16.0 \sim
-10.6$, but rather close to $g_{KN\Sigma}^{}$ which is $3.1 \sim 4.6$
\cite{JRVD01}.
In this work, we use $g_{KN\Theta}^{} = 2.2$ following Ref.
\cite{PSTCG00}.
The only undetermined parameter in Eq.~(\ref{Lag:Theta}) is
$\kappa_\Theta^{}$, the anomalous magnetic moment of the $\Theta^+$,
which should reveal the structure of the $\Theta^+$.
In Ref. \cite{NHK03}, the authors estimated $\kappa_\Theta^{}$ in
several models.
For example, they obtained $\kappa_\Theta^{} \sim -0.7$ in the
diquark-diquark-antiquark picture of Jaffe and Wilczek \cite{JW03},
while $\kappa_\Theta^{} \sim -0.4$ if the $\Theta^+$ is a $KN$ system.
These values are different from the chiral quark soliton model which
gives $\kappa_\Theta^{} \sim +0.3$ \cite{Kim03}.
In this work, we will treat $\kappa_\Theta^{}$ as a free parameter and
give the results in the range of $-0.7 < \kappa_\Theta^{} < +0.7$.

The Lagrangians of Eq.~(\ref{Lag:K*}) contain four coupling constants.
The coupling $g_{K^* K \gamma}$ is estimated from the experimental data
for $K^*$ radiative decays.
The decay width is given by
\begin{equation}
\Gamma_{K^* \to K \gamma} = \frac{g_{K^* K \gamma}^2}{12\pi} |{\bf
p}_\gamma^{}|^3.
\end{equation}
Using the experimental values \cite{PDG02}, we obtain $g_{K^* K
\gamma}^{} = 0.388$ GeV$^{-1}$ for the neutral decay and $g_{K^* K
\gamma}^{} = 0.254$ GeV$^{-1}$ for the charged decay.
However, there is no information on the couplings $g_{K^*N\Theta}^{}$,
$\kappa^T_{K^* n \Theta}$, and $\kappa^T_{K^* p \Theta}$.
Since the decay of the $\Theta^+$ into $K^* N$ is not kinematically
allowed, we have to rely on theoretical estimate which is, however, not
available until now.
The only hint we have is that $g_{K^* N \Lambda}^{} \sim -4.5$ and $g_{K^*
N \Sigma}^{} \sim -2.6$, which are smaller than $g_{K N \Lambda}^{}$ and
$g_{KN\Sigma}^{}$ by a factor of 2.4--3.5 or 1.2--1.8 \cite{WF88}.
{}From this observation, we expect that $g_{K^*N\Theta}^{}$ would be
smaller than $g_{KN\Theta}^{}$.
However, their relative phase is still unfixed.
Thus, we treat $g_{K^*N\Theta}^{}$ as a free parameter and give the
results by varying $g_{K^*N\Theta}^{}$.
We shall find that measuring pion-induced process together with the
photon-induced processes will give us a clue on the $g_{K^*N\Theta}^{}$
coupling.
The tensor couplings $\kappa_{K^* n \Theta}^{T}$ and $\kappa_{K^* p \Theta}^T$
should also be examined, but it will {\em not\/} be considered in this
exploratory study.

The photoproduction amplitudes are in general written as
\begin{equation}
T = \varepsilon_\mu \bar{u}_\Theta^{}(p') \, \mathcal{M}^\mu \, u_N^{}(p),
\end{equation}
where $\varepsilon_\mu$ is the photon polarization vector.
With the effective Lagrangians above, it is straightforward to obtain
the production amplitudes.
For $\gamma n \to K^- \Theta^+$ reaction (Fig.~\ref{fig:gam-n}), we have
\begin{eqnarray}
\mathcal{M}_{1(a)}^\mu &=& -\frac{ie g_{KN\Theta}^{} (2q^\mu -
k^\mu)}{t - M_K^2} \gamma_5 F_{1(a)}(s,t,u),
\nonumber \\
\mathcal{M}_{1(b)}^\mu &=& \frac{g_{K^*K\gamma}^{} g_{K^*N\Theta}^{}}
{t - M_{K^*}^2} \varepsilon^{\mu\nu\alpha\beta} k_\alpha q_\beta
\gamma_\nu F_{1(b)}(s,t,u),
\nonumber \\
\mathcal{M}_{1(c)}^\mu &=& -\frac{e\kappa_n}{2M_N}
\frac{g_{KN\Theta}^{}}{s - M_N^2} \gamma_5 ( k\!\!\!/ + p\!\!\!/ +
M_N ) \sigma^{\mu\nu} k_\nu F_{1(c)}(s,t,u),
\nonumber \\
\mathcal{M}_{1(d)}^\mu &=& \frac{ie g_{KN\Theta}^{}}{u
-M_\Theta^2} \left( \gamma_\mu + \frac{i \kappa_\Theta}{2M_\Theta}
\sigma^{\mu\nu} k_\nu \right) \left( p\!\!\!/ - q\!\!\!/ + M_\Theta
\right) \gamma_5 F_{1(d)}(s,t,u),
\end{eqnarray}
and for $\gamma p \to \bar{K}^0 \Theta^+$ (Fig. \ref{fig:gam-p}) we get
\begin{eqnarray}
\mathcal{M}_{2(a)}^\mu &=& - \frac{g_{K^*K\gamma}^{} g_{K^*N\Theta}^{}}
{t - M_{K^*}^2} \varepsilon^{\mu\nu\alpha\beta} k_\alpha q_\beta
\gamma_\nu F_{2(a)}(s,t,u),
\nonumber \\
\mathcal{M}_{2(b)}^\mu &=& - \frac{ie g_{KN\Theta}^{}}{s
-M_N^2} \gamma_5
\left( k\!\!\!/ + p\!\!\!/ + M_N \right)
\left( \gamma_\mu + \frac{i \kappa_p}{2M_N}
\sigma^{\mu\nu} k_\nu \right)
F_{2(b)}(s,t,u),
\nonumber \\
\mathcal{M}_{2(c)}^\mu &=& - \frac{ie g_{KN\Theta}^{}}{u
-M_\Theta^2} \left( \gamma_\mu + \frac{i \kappa_\Theta}{2M_\Theta}
\sigma^{\mu\nu} k_\nu \right) \left( p\!\!\!/ - q\!\!\!/ + M_\Theta
\right) \gamma_5 F_{2(c)}(s,t,u).
\end{eqnarray}

Here we have multiplied form factors which take into account the
structure of each vertex.
The gauge invariance ($k \cdot \mathcal{M} = 0$) is then easily checked
when all the form factors are set to one.
In $\gamma n \to K^- \Theta^+$, the $t$-channel $K^*$ exchange and the
$s$-channel terms have gauge-invariant forms.
The gauge non-invariant part of the $t$-channel $K$ exchange is canceled
by that of the $u$-channel terms.
In $\gamma p \to \bar{K}^0 \Theta^+$, the $t$-channel $K^*$ exchange
is gauge-invariant.
And the sum of the $s$-channel and $u$-channel is gauge-invariant.
But introducing different form factors at each vertex spoils gauge
invariance.
In Ref. \cite{NHK03}, to keep gauge invariance and to include the
effects of form factors, the cross section for $\gamma n \to K^- \Theta^+$
obtained from tree graphs is multiplied by 0.18.
This factor is the average value of $\sigma_{\rm expt.}/
\sigma_{\rm theory}$ in $\gamma p \to K^+ \Lambda$ reaction near
threshold, where $\sigma_{\rm theory}$ contains the contribution from the
Born terms only.
However, in $\gamma p \to K^+ \Lambda$ there are other production
mechanisms such as nucleon and hyperon resonances which interfere with
the $K$/$K^*$ exchanges and the nucleon Born terms \cite{MB99,JRVD01,HCKC01}.
Thus assuming the same suppression in $\Lambda$ photoproduction and
$\Theta^+$ photoproduction is questionable and should be further tested.
In Refs. \cite{LK03a,LK03b}, the authors used the same form factor for
all channels which is a function of $\sqrt{s}$ only,
\begin{equation}
F(s) = \frac{\Lambda^2}{\Lambda^2 + {\bf q}_i^2},
\label{ff:ko}
\end{equation}
where ${\bf q}_i$ is the three-momentum of the initial state particles in
the CM frame,
\begin{equation}
{\bf q}_i^2 = \frac{1}{4s} \lambda(s,M_N^2,0),
\end{equation}
with $\lambda(x,y,z) = x^2 + y^2 + z^2 - 2(xy+yz+zx)$.
Therefore, as the energy becomes larger, the cross section becomes
smaller since $F(s)$ decreases rapidly with energy.
Since the form of the form factors and the cutoff parameters should be
justified by experimental data, measuring the total and differential cross
sections should discern the form factor dependence of the cross sections.

\begin{figure}[t]
\centering
\epsfig{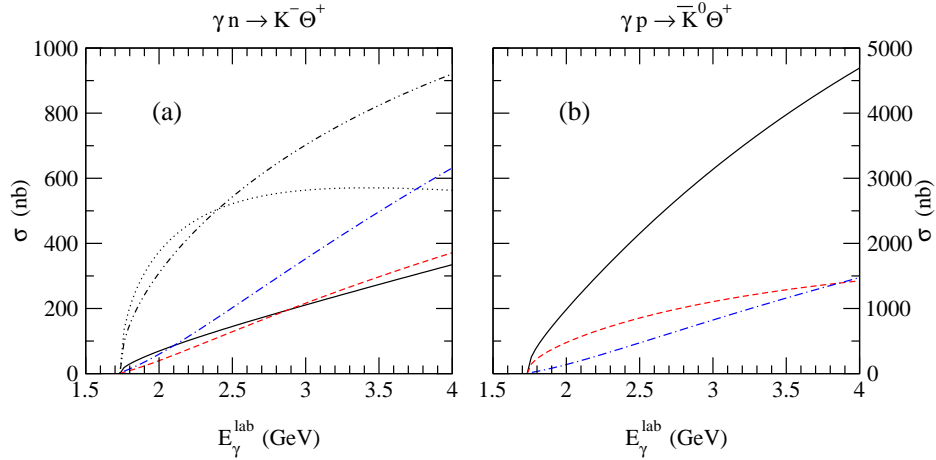}
\caption{Cross sections for (a) $\gamma n \to K^- \Theta^+$ and (b)
$\gamma p \to \bar{K}^0 \Theta^+$ reactions without form factors.
The dashed lines are the results without the $K^*$ exchange and the
dot-dashed lines are from the $K^*$ exchange only. The solid lines are their
sums. In (a), the dotted line is from the ($t+u$)-channel and the
dot-dot-dashed line is the $s$-channel result.}
\label{fig:gam-woff}
\end{figure}

In this work, we take another approach for the form factors.
Motivated by the analyses of kaon photoproduction \cite{JRVD01},
we use the form factor \cite{PJ91}
\begin{equation}
F(r,M_{\rm ex}) = \frac{\Lambda^4}{\Lambda^4 + (r - M_{\rm ex}^2)^2},
\label{ff:ours}
\end{equation}
for {\it each vertex\/}.
Here $M_{\rm ex}$ is the mass of the exchanged particle and $r$ is the
square of the transferred momentum.
This has the correct on-shell condition that $F(r,M_{\rm ex}) = 1$ at
$r=M_{\rm ex}^2$.
However introducing different form factors depending on the channels
breaks gauge invariance.
There are several recipes in restoring gauge invariance with the use of
phenomenological form factors \cite{Ohta89,HBMF98a,DW01a-b}.
But the results depend on the employed form factors and the way to
restore gauge invariance.
In order to avoid any complication and to keep gauge invariance
in a simple way, we use
\begin{eqnarray}
F_{1(a)} = F_{1(d)} = \left\{ F(t,M_K)^2 + F(u,M_\Theta)^2 \right\}/2,
\quad
F_{1(b)} = F(t,M_{K^*})^2, \quad F_{1(c)} = F(s,M_N)^2,
\nonumber \\
\end{eqnarray}
and
\begin{equation}
F_{2(a)} = F(t,M_{K^*})^2, \qquad
F_{2(b)} = F_{2(c)} = \left\{ F(s,M_N)^2 + F(u,M_\Theta)^2 \right\}/2.
\end{equation}
This is an unsatisfactory aspect of this hadronic model approach, but it
should be sufficient for this qualitative study.
For comparison, we will give the results obtained with this form factor
and those with the form factor of Eq.~(\ref{ff:ko}).

\begin{figure}[t]
\centering
\epsfig{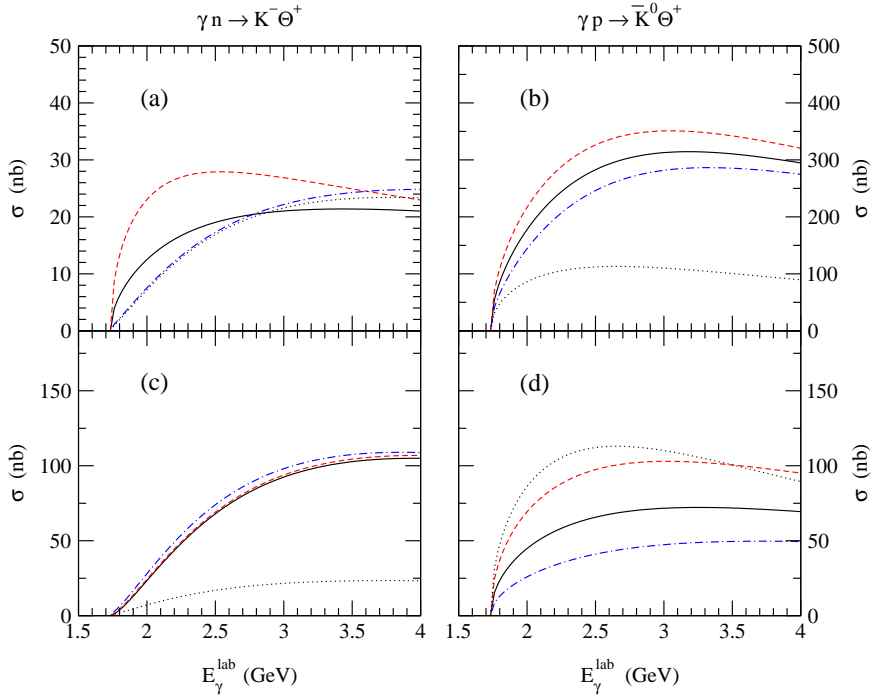}
\caption{Cross sections for (a,c) $\gamma n \to K^- \Theta^+$ and (b,d)
$\gamma p \to \bar{K}^0 \Theta^+$ reactions with the form factors of
Eq.~(\ref{ff:ko}) with $\Lambda = 0.75$ GeV.
In (a,b), $g_{K^*N\Theta}^{} = +2.2$ is used while $g_{K^*N\Theta}^{} = -2.2$
in (c,d). The solid lines are obtained with $\kappa_\Theta^{} = 0$,
while the dashed and dot-dashed lines are with $\kappa_\Theta^{} = +0.7$
and $-0.7$, respectively. The dotted lines correspond to
$g_{K^*N\Theta}^{}=0$.}
\label{fig:gam-ffko}
\end{figure}

We are now ready to calculate the cross sections for $\Theta^+$
photoproduction.
In Fig.~\ref{fig:gam-woff}, the total cross sections for
$\gamma n \to K^- \Theta^+$ and $\gamma p \to \bar{K}^0 \Theta^+$ are
given without form factors.
For this calculation we set $g_{K^*N\Theta}^{} = g_{KN\Theta}^{}$ and
$\kappa_\Theta^{} = 0$.
In Fig.~\ref{fig:gam-woff}, the dashed lines are obtained without the $K^*$
exchanges and the dot-dashed lines are with the $K^*$ exchanges alone.
The solid lines are their sums.
This shows that the cross section for $\gamma p \to \bar{K}^0 \Theta^+$ is
much larger than that for $\gamma n \to K^- \Theta^+$.
This holds even in the absence of the $K^*$ exchanges (dashed lines
in Fig.~\ref{fig:gam-woff}).
Several comments are in order.
In $\gamma n \to K^- \Theta^+$ reaction, there are strong (destructive)
interference among the channels.
In contrast to the assumption of Ref. \cite{CN03}, the contributions
from the $s$- and $u$-channels are not suppressed.
In both reactions, we found that the $K^*$ exchange is not suppressed
compared to the other production mechanisms although it strongly
depends on the unknown coupling $g_{K^*N\Theta}^{}$.
It is interesting to note that the $K^*$ exchange interferes destructively
with the other amplitudes in $\gamma n \to K^- \Theta^+$ but
constructively in $\gamma p \to \bar{K}^0 \Theta^+$.
If the relative phase between $g_{KN\Theta}^{}$ and $g_{K^*N\Theta}^{}$
is changed, then the interference patterns are reversed.
In this case, $\gamma n \to K^- \Theta^+$ has larger cross section than
$\gamma p \to \bar{K}^0 \Theta^+$ when $E_\gamma^{\rm lab} < 2.5$ GeV.
For $E_\gamma^{\rm lab} > 2.5$ GeV, the cross section for
$\gamma p \to \bar{K}^0 \Theta^+$ becomes larger and increases faster
than that of $\gamma n \to K^- \Theta^+$.

\begin{figure}[t]
\centering
\epsfig{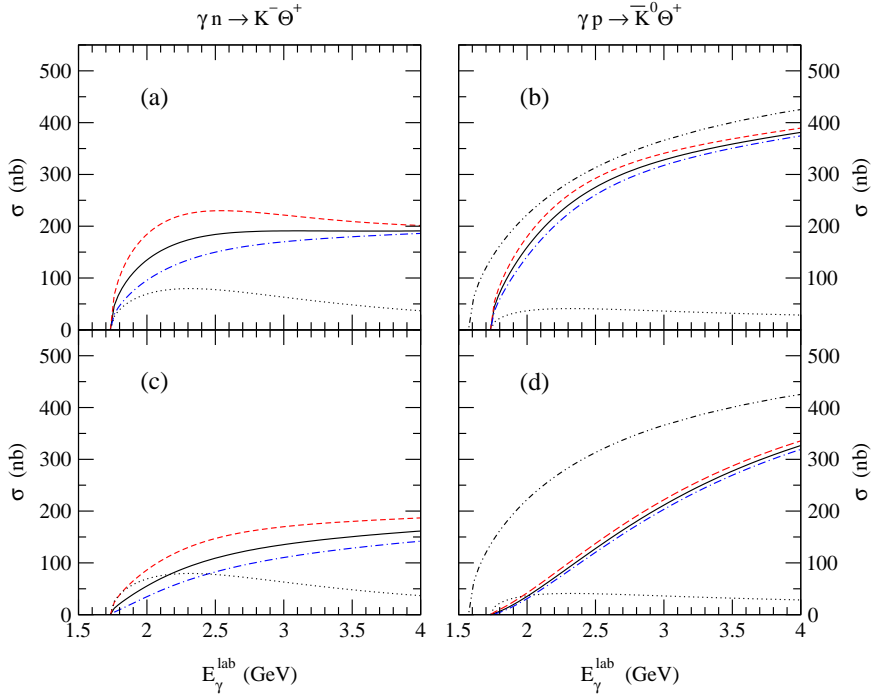}
\caption{Cross sections for (a,c) $\gamma n \to K^- \Theta^+$ and (b,d)
$\gamma p \to \bar{K}^0 \Theta^+$ reactions with the form factors of
Eq.~(\ref{ff:ours}).
The notations are the same as in Fig.~\ref{fig:gam-ffko}. In (b,d) the
total cross section for $\phi$ photoproduction \cite{OTL00} is given by
dot-dot-dashed lines for comparison.}
\label{fig:gam-ff18}
\end{figure}

We now investigate the form factor dependence of the cross sections.
Given in Fig.~\ref{fig:gam-ffko} are the cross sections obtained with
the form factor prescription of Eq.~(\ref{ff:ko}) with $\Lambda = 0.75$
GeV.
The upper graphs are obtained with $g_{K^*N\Theta}^{} = +2.2$ and the lower
ones are with $g_{K^*N\Theta}^{} = -2.2$.
In this figure, we also give the results by varying $\kappa_\Theta^{}$
from $-0.7$ to $+0.7$.
The energy dependence of the cross sections are similar for both the
neutron and proton targets, which is expected from the form of
the form factor (\ref{ff:ko}).
When $g_{K^*N\Theta}^{} = +2.2$, the cross section for $\gamma p \to
\bar{K}^0 \Theta^+$ is larger than that for $\gamma n \to K^- \Theta^+$
by a factor of 10.
However when $g_{K^*N\Theta}^{} = -2.2$, the cross section for
$\gamma n \to K^- \Theta^+$ is slightly larger.
In order to see the contributions from the $K^*$ exchange, we give the
results without the $K^*$ exchange with $\kappa_\Theta^{}=0$
(dotted lines in Fig.~\ref{fig:gam-ffko}).

In Fig.~\ref{fig:gam-ff18}, we present our results with the form factors of
Eq.~(\ref{ff:ours}).
Here the results without the $K^*$ exchange and with $\kappa_\Theta^{}=0$
are also shown by the dotted lines.
The upper graphs are obtained with $g_{K^*N\Theta}^{} = +2.2$ and the lower
ones with $g_{K^*N\Theta}^{} = -2.2$.
In this calculation we use the cutoff
\begin{equation}
\Lambda = 1.8 \mbox{ GeV},
\end{equation}
as fixed in the study of $\Lambda$ photoproduction with the same form
factor prescription \cite{JRVD01}.
We see that with this form factor the final results are not so sensitive
to the phase of the $g_{K^*N\Theta}^{}$ coupling, indicating the large
contributions from the $K^*$ exchanges.
We notice that our results are consistent with the observation of SAPHIR
Collaboration that the cross section for $\gamma p \to \bar{K}^0 \Theta^+$
is about 200 nb in the photon energy range from 1.7 to 2.6 GeV and
it is similar to the cross section for $\phi$ photoproduction.
For comparison, we give the total cross section for $\phi$
photoproduction in the literature \cite{OTL00} by the dot-dot-dashed lines
in Figs.~\ref{fig:gam-ff18}(b,d).
This shows that the positive phase of $g_{K^*N\Theta}^{}$ is favored by
SAPHIR experiment.
In addition, the cross section for $\gamma p \to \bar{K}^0 \Theta^+$
is larger in most cases.
We also find that the energy dependence of the cross sections
for the two reactions are different.
In $\gamma n \to K^- \Theta^+$, the cross section saturates as the energy
increases, while it keeps increasing with the energy in
$\gamma p \to \bar{K}^0 \Theta^+$.
The $\kappa_\Theta^{}$ dependence is negligible for the photon-proton
reaction, while in photon-neutron reaction the cross section nontrivially
depends on $\kappa_\Theta^{}$.
Since different choice of the form factors gives different results, as
can be seen in Figs.~\ref{fig:gam-ffko} and \ref{fig:gam-ff18},
it would be very useful to have measurements on the total and especially
differential cross sections to justify a particular type of the form
factors and to discriminate the role of the each production channel.

\section{$\bm{\pi^- p \to K^- \Theta^+}$}

\begin{figure}[t]
\centering
\epsfig{file=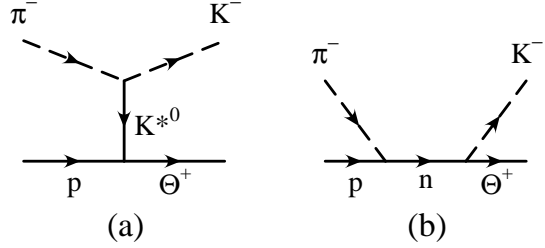, width=0.45\hsize}
\caption{Diagrams for $\pi^- p \to K^- \Theta^+$ reaction.}
\label{fig:pi-p}
\end{figure}

We now turn to the pion-induced reaction, $\pi^- p \to K^- \Theta^+$, of
which tree diagrams are shown in Fig.~\ref{fig:pi-p}.
Without the $K^*$ exchange, only the $s$-channel diagram is allowed
which was considered in Ref. \cite{LK03a}.
The possible $u$-channel diagram includes $\Theta'$ whose minimal quark
content should be $uuud\bar{s}$.
Such an exotic baryon may be a member of the isotensor pentaquarks as suggested
by Ref. \cite{CPR03}, but its existence seems to be disfavored by SAPHIR
experiment \cite{SAPHIR03}.
Thus we do not consider the $u$-channel diagrams in this study.
Note also that the $s$-channel diagrams containing the nucleon
resonances are neglected as in the $\Theta^+$ photoproduction study
of the previous section.
Furthermore, we will find that the $t$-channel $K^*$ exchange is
the most dominant process.
For this calculation, in addition to $\mathcal{L}_{K^*N\Theta}$ of
Eq.~(\ref{Lag:K*}), we need the effective Lagrangian $\mathcal{L}_{K^*
K\pi}$, which reads
\begin{equation}
\mathcal{L}_{K^*K\pi} = -ig_{K^*K\pi}^{} \left\{ \bar{K} \partial^\mu
\pi K_\mu^* - \partial^\mu \bar{K} \pi K_\mu^* \right\} + \mbox{ h.c.},
\end{equation}
where $K^T = (K^+, K^0)$, $\bar{K} = (K^-, \bar{K}^0)$, etc, and
\begin{equation}
\mathcal{L}_{\pi NN} = -i g_{\pi NN}^{} \bar{N} \gamma_5 \pi N,
\end{equation}
with $\pi = \bm{\pi} \cdot \bm{\tau}$.
The coupling constant $g_{K^*K\pi}^{}$ fixed by the experimental data
for $K^* \to K \pi$ decay,
\begin{equation}
\Gamma_{K^* \to K\pi} = \frac{g_{K^*K\pi}^2}{2\pi M_{K^*}^2} |{\bf p}_\pi|^3,
\end{equation}
is $g_{K^*K\pi} = 3.28$ which is comparable to the SU(3)
symmetry value, 3.02.
We also use $g_{\pi NN}^2/(4\pi) = 14$.

The production amplitude of this reaction is written as
\begin{equation}
T = \bar{u}_\Theta^{}(p') \, \mathcal{M} \, u_p(p),
\end{equation}
where the diagrams of Fig.~\ref{fig:pi-p} give
\begin{eqnarray}
\mathcal{M}_{6(a)} &=& \frac{\sqrt2 g_{K^*K\pi}^{} g_{K^*N\Theta}^{}}
{t - M_{K^*}^2} \left\{ k\!\!\!/ + q\!\!\!/ + \frac{M_K^2-
M_\pi^2}{M_{K^*}^2} (k\!\!\!/ - q\!\!\!/) \right\} F_{6(a)}(s,t,u),
\nonumber \\
\mathcal{M}_{6(b)} &=& \frac{\sqrt2 g_{KN\Theta}^{} g_{\pi NN}^{}}
{s - M_N^2} \left( k\!\!\!/ + p\!\!\!/ - M_N \right)
F_{6(b)}(s,t,u).
\end{eqnarray}
As in photoproduction, we work with two choices of the form factors.
First, following Ref. \cite{LK03a}, we set
\begin{equation}
F_{6(a)}(s,t,u) = F_{6(b)}(s,t,u) = \frac{\Lambda^2}{\Lambda^2 +
{\bf q}_i^2},
\label{ff:ko2}
\end{equation}
as in Eq.~(\ref{ff:ko}) with ${\bf q}_i^2 = \lambda(s,M_N^2,M_\pi^2)/4s$.
We also employ the covariant form factors as
\begin{equation}
F_{6(a)}(s,t,u) = F(t,M_{K^*})^2, \qquad F_{6(b)}(s,t,u) = F(s,M_N)^2,
\label{ff:ours2}
\end{equation}
where $F(r,M)$ is given in Eq.~(\ref{ff:ours}).

\begin{figure}[t]
\centering
\epsfig{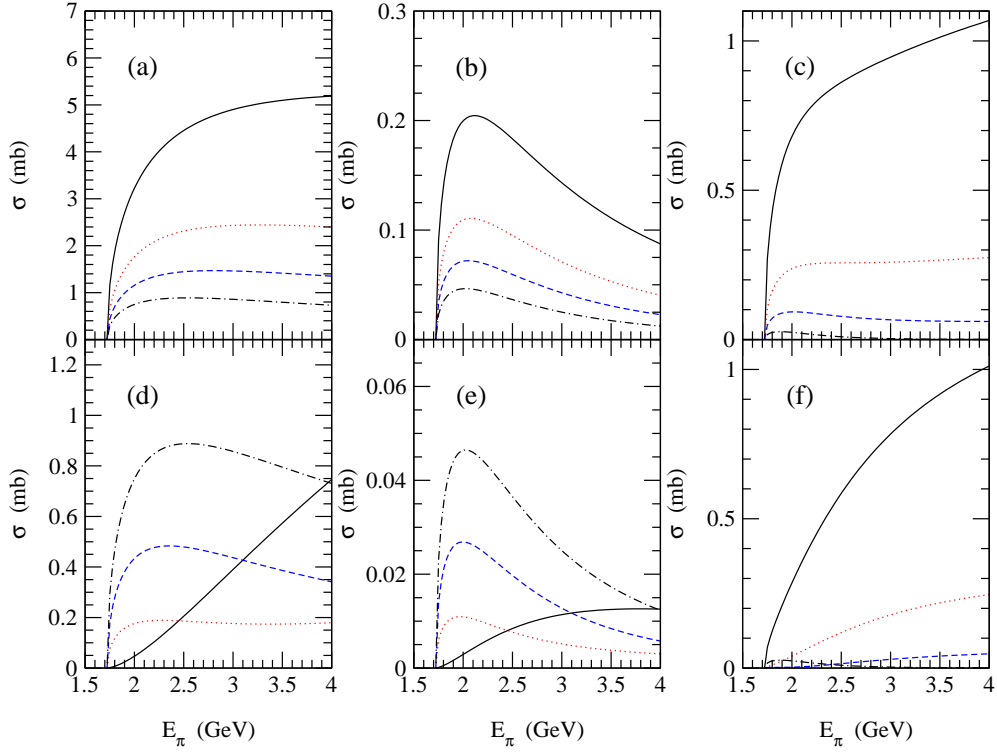}
\caption{Cross sections for $\pi^- p \to K^- \Theta^+$
(a,d) without form factors, (b,e) with the form factors of Eq.~(\ref{ff:ko})
with $\Lambda = 0.5$ GeV, (c,f) with the form factors of
Eq.~(\ref{ff:ours}) with $\Lambda = 1.8$ GeV.
In (a,b,c), the solid, dotted, dashed and dot-dashed lines are with
$g_{K^*N\Theta}^{} = -2.2$, $-1.1$, $-0.5$, and $0.0$, respectively. 
In (d,e,f), the solid, dotted, dashed and dot-dashed lines are with
$g_{K^*N\Theta}^{} = 2.2$, $1.1$, $0.5$, and $0.0$, respectively.}
\label{fig:pi-tot}
\end{figure}

The results are shown in Fig.~\ref{fig:pi-tot}.
In Fig.~\ref{fig:pi-tot}(a), we show the results without form factors
with $g_{K^*N\Theta}^{}<0$, where the $t$ and $s$-channels interfere
constructively.
Given in Fig.~\ref{fig:pi-tot}(d) are the results with
$g_{K^*N\Theta}^{}>0$, where we have destructive interference.
In these plots, we give the results by varying the value of
$g_{K^*N\Theta}^{}$.
This evidently shows an important role driven by the $t$-channel $K^*$
exchange.
We see that, even with $g_{K^*N\Theta}^{}=\pm 0.5$,
$K^*$ exchange cannot be neglected.
This can be seen by comparing with the results of the $s$-channel alone
(the dot-dashed lines).
In Fig.~\ref{fig:pi-tot}(b,e), we show the results with the form factor
of Eq.~(\ref{ff:ko2}). 
Following Ref. \cite{LK03a}, we use $\Lambda = 0.5$ GeV.
Here again the role of $K^*$ exchange can be easily found.
In this reaction, we also found that the form factor (\ref{ff:ko2})
suppresses the cross section by an order of magnitude, which is
due to the soft cutoff value.
If we use $\Lambda = 0.75$ GeV in Eq.~(\ref{ff:ko2}), the peak value in
Fig.~\ref{fig:pi-tot}(b) would be 0.6 mb.
The effect of the $K^*$ exchange is even more drastic when we use the
form factor of Eq.~(\ref{ff:ours2}).
Figure~\ref{fig:pi-tot}(c,f) show our results with the form factor
of Eq.~(\ref{ff:ours2}) and $\Lambda = 1.8$ GeV.
Here the results without the $K^*$ exchange is very much suppressed with
the peak value being only about 25 $\mu$b.
But with the $K^*$ exchange the cross sections become a few hundred $\mu$b
depending on the magnitude of $g_{K^*N\Theta}^{}$.
We also found that the cross sections are not so sensitive to the phase
of $g_{K^*N\Theta}^{}$, thus $\pi^- p \to K^- \Theta^+$ reaction can be
a good place to measure its magnitude.

\section{Production of negative-parity $\bm{\Theta^+}$}

We now consider the production of the $\Theta^+$ assuming that it has
negative parity.
This process was considered in Ref. \cite{NHK03} for photon-neutron
reaction and found to have smaller cross sections than the
positive-parity $\Theta^+$ production.
In this case, the effective Lagrangians in Eqs. (\ref{Lag:Theta}) and
(\ref{Lag:K*}) are changed as
\begin{eqnarray}
\mathcal{L}_{KN\Theta} &=& g_{KN\Theta}^{} ( \bar{\Theta} K^+ n -
\bar{\Theta} K^0 p ) + \mbox{ h.c.},
\nonumber \\
\mathcal{L}_{K^* N \Theta} &=& -i g_{K^*N\Theta}^{} (
\bar{\Theta} \gamma_5 \gamma^\mu K^{*+}_\mu n
-  \bar{\Theta} \gamma_5 \gamma^\mu K^{*0}_\mu p)
+ \mbox{ h.c.},
\end{eqnarray}
where we have dropped the tensor coupling terms of the $K^*N\Theta$
interaction as its effects will not be considered throughout this work.
The form of the above effective Lagrangians is obtained with the
prescription of Ref. \cite{BMZ95}, which was used to construct the effective
Lagrangians for the interactions of the $J^P = \frac12^-$ $S_{11}(1535)$
resonance.
Here we inserted $i\gamma_5$ in the appropriate place in order to
account for the odd parity of the $\Theta^+$.
Then the decay width of $\Theta^+ \to KN$ becomes
\begin{equation}
\Gamma_{\Theta^+ \to K^+ n + K^0 p}
= \frac{g_{KN\Theta}^2}{2\pi} \frac{|{\bf p}_{K}^{}|
( \sqrt{M_N^2 + {\bf p}_K^2}+M_N)}{M_\Theta},
\label{odd-decay}
\end{equation}
which gives $g_{KN\Theta}^{} = 0.307$ with $\Gamma_\Theta = 5$ MeV.

The production amplitudes for $\gamma n \to K^- \Theta^+$ are
\begin{eqnarray}
\mathcal{M}_{1(a)}^\mu &=& \frac{e g_{KN\Theta}^{} (2q^\mu -
k^\mu)}{t - M_K^2} F_{1(a)}(s,t,u),
\nonumber \\
\mathcal{M}_{1(b)}^\mu &=& \frac{ig_{K^*K\gamma}^{} g_{K^*N\Theta}^{}}
{t - M_{K^*}^2} \varepsilon^{\mu\nu\alpha\beta} k_\alpha q_\beta
\gamma_5 \gamma_\nu F_{1(b)}(s,t,u),
\nonumber \\
\mathcal{M}_{1(c)}^\mu &=& -\frac{ie\kappa_n}{2M_N}
\frac{g_{KN\Theta}^{}}{s - M_N^2} ( k\!\!\!/ + p\!\!\!/ +
M_N ) \sigma^{\mu\nu} k_\nu F_{1(c)}(s,t,u),
\nonumber \\
\mathcal{M}_{1(d)}^\mu &=& -\frac{e g_{KN\Theta}^{}}{u
-M_\Theta^2} \left( \gamma_\mu + \frac{i \kappa_\Theta}{2M_\Theta}
\sigma^{\mu\nu} k_\nu \right) \left( p\!\!\!/ - q\!\!\!/ + M_\Theta
\right) F_{1(d)}(s,t,u).
\end{eqnarray}
For $\gamma p \to \bar{K}^0 \Theta^+$ we have
\begin{eqnarray}
\mathcal{M}_{2(a)}^\mu &=& - \frac{ig_{K^*K\gamma}^{} g_{K^*N\Theta}^{}}
{t - M_{K^*}^2} \varepsilon^{\mu\nu\alpha\beta} k_\alpha q_\beta
\gamma_5 \gamma_\nu F_{2(a)}(s,t,u),
\nonumber \\
\mathcal{M}_{2(b)}^\mu &=& \frac{e g_{KN\Theta}^{}}{s
-M_N^2} 
\left( k\!\!\!/ + p\!\!\!/ + M_N \right)
\left( \gamma_\mu + \frac{i \kappa_p}{2M_N}
\sigma^{\mu\nu} k_\nu \right)
F_{2(b)}(s,t,u),
\nonumber \\
\mathcal{M}_{2(c)}^\mu &=& \frac{e g_{KN\Theta}^{}}{u
-M_\Theta^2} \left( \gamma_\mu + \frac{i \kappa_\Theta}{2M_\Theta}
\sigma^{\mu\nu} k_\nu \right) \left( p\!\!\!/ - q\!\!\!/ + M_\Theta
\right) F_{2(c)}(s,t,u),
\end{eqnarray}
and the $\pi^- p \to K^- \Theta^+$ process has
\begin{eqnarray}
\mathcal{M}_{6(a)} &=& -\frac{i\sqrt2 g_{K^*K\pi}^{} g_{K^*N\Theta}^{}}
{t - M_{K^*}^2} \left\{ k\!\!\!/ + q\!\!\!/ + \frac{M_K^2-
M_\pi^2}{M_{K^*}^2} (k\!\!\!/ - q\!\!\!/) \right\} \gamma_5 F_{6(a)}(s,t,u),
\nonumber \\
\mathcal{M}_{6(b)} &=& -\frac{i\sqrt2 g_{KN\Theta}^{} g_{\pi NN}^{}}
{s - M_N^2} \left( k\!\!\!/ + p\!\!\!/ + M_N \right) \gamma_5
F_{6(b)}(s,t,u).
\end{eqnarray}

\begin{figure}[t]
\centering
\epsfig{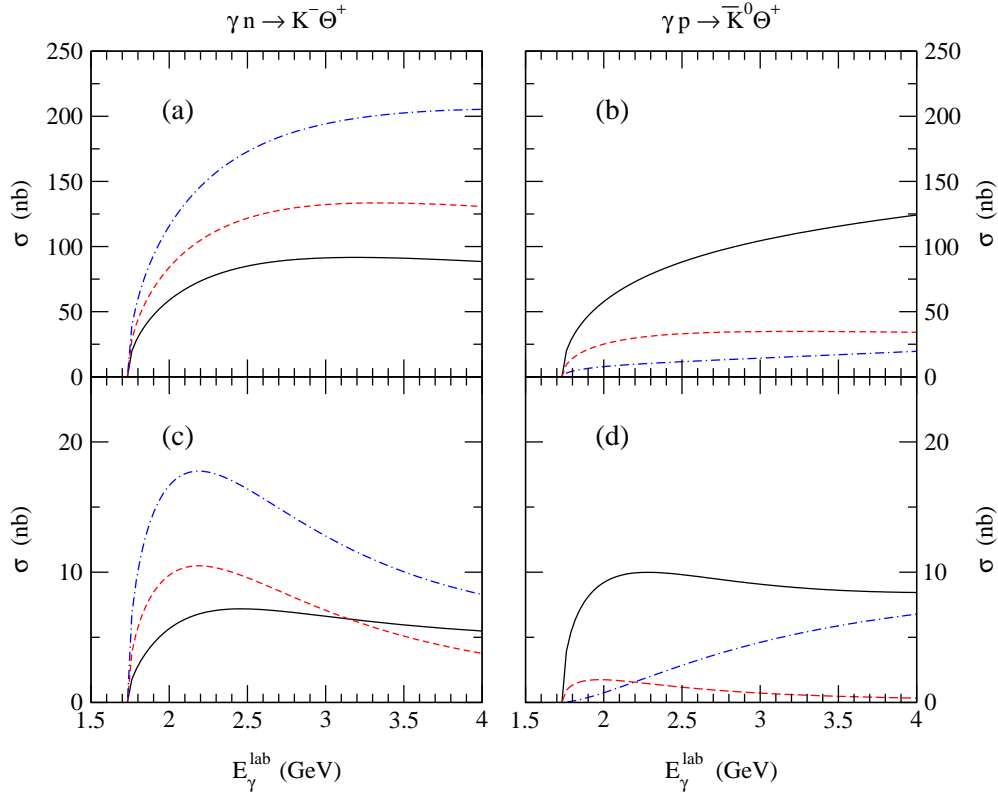}
\caption{Cross sections for (a,c) $\gamma n \to K^- \Theta^+$ and (b,d)
$\gamma p \to \bar{K}^0 \Theta^+$ when the $\Theta^+$ has negative-parity.
In (a,b) no form factors are used and in (c,d) the form factors
(\ref{ff:ours}) are used with $\Lambda=1.8$ GeV.
The solid lines are the results with $g_{K^*N\Theta}^{} = g_{KN\Theta}^{}$,
the dashed lines with $g_{K^*N\Theta}^{} = 0$, and
the dot-dashed lines with $g_{K^*N\Theta}^{} = -g_{KN\Theta}^{}$.}
\label{fig:odd-gam}
\end{figure}

\begin{figure}[t]
\centering
\epsfig{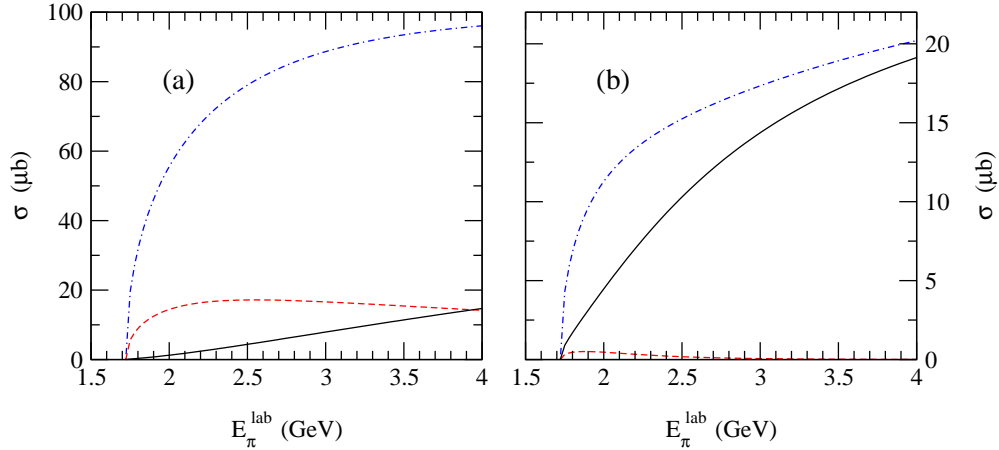}
\caption{Cross sections for $\pi^- p \to K^- \Theta^+$
when the $\Theta^+$ has negative-parity.
In (a) no form factors are used and in (b) the form factors
(\ref{ff:ours}) are used with $\Lambda=1.8$ GeV.
The notations are the same as in Fig.~\ref{fig:odd-gam}.}
\label{fig:odd-pi}
\end{figure}

The results are given in Figs.~\ref{fig:odd-gam} and \ref{fig:odd-pi}
with $g_{KN\Theta}^{} = 0.307$.
Given in Fig.~\ref{fig:odd-gam} are the photon-neutron and photon-proton
reaction results and the pion-proton reaction results are shown in
Fig.~\ref{fig:odd-pi}.
In these figures the solid lines are obtained with $g_{K^*N\Theta}^{} =
g_{KN\Theta}^{}$, the dashed lines are with $g_{K^*N\Theta}^{} = 0$, and
the dot-dashed lines are with $g_{K^*N\Theta}^{} = - g_{KN\Theta}^{}$.
Here we set $\kappa_\Theta = 0$.
In Ref.~\cite{NHK03}, it is claimed that the cross section for the
negative-parity $\Theta^+$ is suppressed compared with the
positive-parity $\Theta^+$ in the case of photon-neutron reaction.
We confirmed this conclusion and found that this is true not only in
photon-neutron reaction, but also in photon-proton and pion-proton
reactions.
In photoproduction, the cross sections are small, less than 200 nb
even without form factors, and the cross section for the photon-neutron
reaction is slightly larger than that for the photon-proton reaction.
If the form factors of Eq.~(\ref{ff:ours}) are used with $\Lambda = 1.8$
GeV, the cross sections become less than 20 nb.
Therefore this seems to be inconsistent with the SAPHIR observation that
the cross section for photon-proton reaction is about 200 nb.
The suppression of the cross section is also seen in the case of
pion-proton reaction (Fig.~\ref{fig:odd-pi}).
In this reaction, the cross section for the negative-parity $\Theta^+$
is less than 100 $\mu$b without form factors and less than 20 $\mu$b with
the form factors.
Thus precise measurements on the production processes may distinguish
the parity of the $\Theta^+$ baryon.

\section{Summary and Discussion}

We have investigated the production processes of the exotic $\Theta^+$
which was discovered recently by several experiments.
We considered photon-induced production reactions, which were used
in the experiments of the LEPS Collaboration, CLAS Collaboration, and
SAPHIR Collaboration.
In addition, we have considered pion-induced production which is
used by the current KEK experiments and is also available at GSI.

Previous analyses on these reactions \cite{LK03a,LK03b,NHK03} are
improved and extended by including some missing interactions and $K^*$
exchanges.
We also employ the form factors and cutoff parameters motivated by
the study of $\Lambda$ photoproduction.
Our results show that the photon-proton reaction cross section is
somehow larger than that of the photon-neutron reaction.
This is in contrast to the conclusion of Refs. \cite{LK03a,LK03b}, which
predicted larger cross section for the photon-neutron reaction.
This is mainly because the model of Refs. \cite{LK03a,LK03b} does not
include the tensor coupling of the electromagnetic interactions and the
$K^*$ exchanges were not coherently included in both reactions.
The cross sections of the two reactions, photon-neutron and photon-proton
reactions, are in the range of a few hundred nb, which seems to be consistent
with the SAPHIR observation as far as photon-proton reaction is
concerned.
But they have different energy dependence which should be verified by
experiments and can test the models adopted in this study.
Furthermore, SAPHIR Collaboration claimed that the cross section of
$\gamma p \to \bar{K}^0 \Theta^+$ is similar to that of $\gamma p \to \phi p$.
If this is confirmed, the positive phase of the $g_{K^*N\Theta}^{}$ seems
to be favored. [See Fig.~\ref{fig:gam-ff18}(b).]

The pion-induced reaction has much larger cross section than the
photon-induced reactions.
We found that this reaction is very sensitive to the magnitude of the
$K^*N\Theta$ coupling but not so to its phase.
So measurement of this reaction would give us a guide to estimate the
magnitude of $g_{K^*N\Theta}^{}$.
Due to lack of information, we could not investigate the role of the tensor
couplings of the $K^*N\Theta$ interaction and the role of nucleon
resonances like the nucleon analog of the $\Theta^+$ in the $s$-channels.
Therefore, theoretical studies on the $\Theta$ couplings in various
models are highly desirable and useful to understand the structure and
the production mechanisms of the $\Theta^+$.

\begin{figure}[t]
\centering
\epsfig{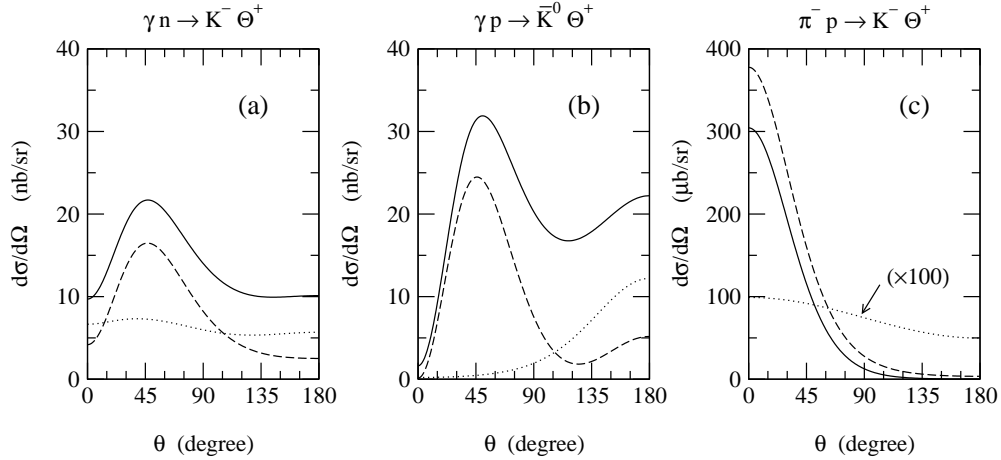}
\caption{Differential cross sections for (a) $\gamma n \to K^- \Theta^+$
at $E_\gamma^{\rm lab} = 2.5$ GeV, (b) $\gamma p \to \bar{K}^0 \Theta^+$
at $E_\gamma^{\rm lab} = 2.5$ GeV, and (c) $\pi^- p \to K^- \Theta^+$ at
$E_\pi^{\rm lab} = 2.5$ GeV, when the $\Theta^+$ has positive-parity.
The form factors (\ref{ff:ours}) are used with $\Lambda=1.8$ GeV.
The solid, dotted, and dashed lines are obtained with $g_{K^*N\Theta}^{} =
g_{KN\Theta}^{}$, $g_{K^*N\Theta}^{} = 0$, and $g_{K^*N\Theta}^{}
= -g_{KN\Theta}^{}$, respectively, where $g_{KN\Theta}^{} = 2.2$.
In (c), the dotted line is exaggerated by a factor of 100.}
\label{fig:diffcs}
\end{figure}

Since we are employing covariant form factors which are different from
the form factors used in previous studies \cite{LK03a,LK03b,NHK03},
it will be useful to measure differential cross sections, which will
discriminate between different form factors.
In Fig.~\ref{fig:diffcs}, we give our predictions on the differential
cross sections for $\gamma n \to K^- \Theta^+$, $\gamma p \to \bar{K}^0
\Theta^+$, and $\pi^- p \to K^- \Theta^+$ with the photon or pion energy
at 2.5 GeV, where the scattering angle $\theta$ is defined by the
directions of the initial photon (pion) momentum and the final $K$ meson
momentum in the CM frame.
The solid, dotted, and dashed lines are obtained with $g_{K^*N\Theta}^{} =
g_{KN\Theta}^{}$, $g_{K^*N\Theta}^{} = 0$, and $g_{K^*N\Theta}^{}
= -g_{KN\Theta}^{}$, respectively, where $g_{KN\Theta}^{} = 2.2$.
This shows that the differential cross section for $\pi^- p \to K^- \Theta^+$
has forward peak which is due to the dominance of the $K^*$ exchange.
In photoproduction, the differential cross section is suppressed at
forward angles and the peak is at $ \theta \sim 45^\circ$.
In particular, a backward (secondary) peak is observed in the case of
photon-proton reaction.

Finally, we have also investigated the production processes of negative-parity
$\Theta^+$ since several studies claim that the $\Theta^+$ has
odd-parity.
We found that the cross sections in this case are very much suppressed
compared with the positive-parity $\Theta^+$ case.
Thus the interpretation of the $\Theta^+$ as an odd-parity pentaquark
state seems to be disfavored by the SAPHIR observation which claims that
the cross section for photon-proton reaction is about 200 nb.
So if the SAPHIR results are confirmed by other experiments, the
interpretation of the $\Theta^+$ as an $I=0$ and $J^P=\frac12^+$
pentaquark will be strongly supported.
Most importantly, further experimental measurements are crucial to make
any decisive conclusion on the couplings and the structure of the $\Theta^+$.

\acknowledgments

We are grateful to J. Barth, N. I. Kochelev, T. Nakano, and T.-S. H. Lee
for useful information and fruitful discussions.
This work was supported by Korea Research Foundation Grant
(KRF-2002-015-CP0074).

\end{document}